\documentclass[10pt,journal,compsoc]{IEEEtran}
\ifCLASSOPTIONcompsoc
  \usepackage[nocompress]{cite}
\else
  \usepackage{cite}
\fi
\usepackage{amssymb}

\ifCLASSINFOpdf
   \usepackage[pdftex]{graphicx}
\else
\fi
\usepackage{amsmath}
\usepackage{gensymb}
\usepackage{colortbl}
\usepackage{svg}
\usepackage{multirow}
\usepackage{acronym}
\newacro{CNN}[CNN]{Convolutional Neural Network}
\newacro{BoCF}[BoCF]{Bag of Color Features}
\newacro{FC4}[FC$^4$]{Fully Convolutional Color Constancy}
\newacro{MCDE}[MCDE]{Monte Carlo Dropout Ensembles}
\newacro{RAE}[RAE]{recovery angular error}
\newacro{CWCC}[CWCC]{Channel-Wise Color Constancy}
\makeatletter
\renewcommand\@makefntext[1]{\leftskip=2em\hskip-2em\@makefnmark#1}
\makeatother

\begin{document}
\title{Robust channel-wise illumination estimation}


\author{Firas~Laakom,
Jenni~Raitoharju,
Jarno~Nikkanen,
Alexandros~Iosifidis, and~Moncef~Gabbouj

       \thanks{F. Laakom  and M. Gabbouj are with Department of Computing Sciences, Tampere University, Tampere, Finland, Tampere University, Tampere, Finland.}
       \thanks{J. Raitoharju is with the Programme for Environmental Information, Finnish Environment Institute, Jyväskylä, Finland.}
       \thanks{A. Iosifidis is with the Department of Electrical and Computer Engineering, Aarhus University, Aarhus, Denmark.}
       \thanks{J. Nikkanen is with Xiaomi Finland Oy, Tampere, Finland. }
}

\IEEEtitleabstractindextext{%
\begin{abstract}
Recently, Convolutional Neural Networks (CNNs) have been widely used to solve the illuminant estimation problem and have often led to state-of-the-art results. Standard approaches operate directly on the input image. In this paper, we argue that this problem can be decomposed into three channel-wise independent and symmetric sub-problems and propose a novel CNN-based illumination estimation approach based on this decomposition. The proposed method substantially reduces the number of parameters needed  to solve the task while achieving competitive experimental results compared to state-of-the-art methods. Furthermore, the practical application of illumination estimation techniques typically requires identifying the extreme error cases. This can be achieved using an uncertainty estimation technique. In this work, we propose a novel color constancy uncertainty estimation approach that augments the trained model with an auxiliary branch which learns to predict the error based on the feature representation. Intuitively, the model learns which feature combinations are robust and are thus likely to yield low errors and which combinations result in erroneous estimates. We test this approach on the proposed method and show that it can indeed be used to avoid several extreme error cases and, thus, improves the practicality of the proposed technique.
\end{abstract}

\begin{IEEEkeywords}
Color constancy, illumination estimation, robustness, uncertainty estimation
\end{IEEEkeywords}}

\maketitle

\IEEEdisplaynontitleabstractindextext

% \fi
\IEEEpeerreviewmaketitle

%-------------------------------------------------------------------------
\IEEEraisesectionheading{\section{Introduction}\label{sec:intro}}
\IEEEPARstart{T}{he} human visual system is able to adapt to different lighting conditions to produce invariant representations of the objects \cite{ebner2007color}.  This ability to remove the illumination effect on the colors is known as color constancy. Digital cameras try to mimic this ability in their preprocessing pipelines and try to suppress the light source effect on the colors presented in the scene. The central objective is to recover the true colors of the objects observed as if the light source is a neutral illumination. This task in modern cameras is known
as the computational color constancy and several unsupervised \cite{xie2006estimating, qian2019finding,yang2015efficient,banic2017unsupervised,d4,d7,d8}  and supervised approaches \cite{46440,yuan2007color,lou2015color,22,44,mine,f1,Barron2015ConvolutionalCC} have been proposed to solve it.  Achieving an invariant representation of the objects regardless of the illuminant is critical for many other vision tasks such as classification \cite{goodfellow2016deep,rawat2017deep} and scene understanding \cite{nadeem2019deep,yang2018scene,ess2009segmentation,yang2018scene}.

Formally, RGB values of an image I at every pixel  $(x,y)$ are expressed as a function of the illuminant $ \textbf{e}(x,y, \lambda)$, the surface reflectance $\textbf{R}(x,y,\lambda)$, and the camera sensitivity $\textbf{S}(\lambda)$ as follows:
\begin{equation}
            \textbf{I}(x,y) =  \int_\lambda \textbf{e}(x,y, \lambda) \textbf{R}(x,y,\lambda) \textbf{S}(\lambda) d \lambda, 
\end{equation}
where $\lambda$ is the wavelength. Computational color constancy approaches simplify the problem by assuming a global uniform illuminant in the whole scene:
\begin{equation}
       \textbf{e} = \textbf{e}(x,y) =  \int_\lambda \textbf{e}(x,y, \lambda) \textbf{S}(\lambda) d \lambda.
\end{equation}
This leads to the following final equation:
\begin{equation} \label{eq1}
 \textbf{I}(x,y) =  \textbf{R}(x,y)  \circ  \textbf{e},
\end{equation}
where $\circ$ is element-wise multiplication. Based on this equation, computational color constancy is typically achieved in a two-step process. In the first step, the global illuminant  $\textbf{e}$ is estimated. Then, in the second step, the original colors $\textbf{R}$ are restored by pixel-wise normalization of the raw image $ \textbf{I}$ by the estimated $\textbf{e}$. As the second step is a straightforward transformation, the computational color constancy problem is reduced to the first step, i.e., illuminant  estimation. It can be seen that the latter is an ill-posed problem as it has one known variable  $\textbf{I}$ and  two unknowns, $\textbf{e}$ and $\textbf{R}$. For example, given a yellowish pixel, it is impossible to know if it is truly a yellow object under white illuminant  or a white object under yellow illuminant. 

Recently, \ac{CNN}-based approaches have been extensively used to solve this problem  \cite{22,44,laakom2020monte,HoldGeoffroy2017DeepOI},  given their strong abilities to generalize and to regress directly from the input raw image to the desired target variable without needing feature extraction or preprocessing. The main problem in using the \ac{CNN}-based techniques follows from the lack large datasets as even the largest publicly available datasets contain only about 7000 images \cite{laakom2019intel}.
The models can be categorized either as patch-wise or single-pass methods. 
Patch-wise approaches solve the data scarcity problem by training on small patches of the original image \cite{22, mine,bianco2017single,bianco2012color,bianco2014adaptive}.  In the test phase, the patch estimates are aggregated directly using the average  \cite{22} or using a weighted combination \cite{bianco2017single} to obtain the final estimate. Various other patch-based approaches using different combination techniques have been proposed \cite{bianco2012color,bianco2014adaptive}. Due to the limited amount of labeled data, an unsupervised pretraining phase of an autoencoder using auxiliary data was proposed in \cite{mine}.   The single-pass CNN models are trained using the full image as input to estimate the illuminant  \cite{Juarez_2020_CVPR,44,das2018color}. Different methods have been proposed in the literature also in this category. Some of them rely on pretrained classic \ac{CNN} architectures, such as VGG16, to overcome the limited number of training samples \cite{lou2015color,44}. A GAN-based approach was proposed in \cite{das2018color} and a \ac{BoCF} approach that discards the spatial information as it is not important in the color constancy context was proposed in \cite{f1}.

Training large \acp{CNN} requires a large amount of data which is not available in the current illuminant  estimation datasets. Moreover,  \acp{CNN} are typically over-parameterized and, thus, expensive computationally and in terms of energy and time which restricts their usage in low computational power devices such as mobile phones. Therefore, reducing the number of parameters is critical for a deployable color constancy model. Most \ac{CNN}-based illuminant  estimation approaches \cite{22,44,HoldGeoffroy2017DeepOI}  operate directly on the input image $\textbf{I}$ without exploiting the specificities and characteristics of Eq.~\eqref{eq1} defining the problem. We argue that this is not optimal and show that the problem presented in Eq.~\eqref{eq1} can be decomposed channel-wise into three symmetric independent sub-problems.  Based on this decomposition, we propose a novel CNN-based computational color constancy approach, named \ac{CWCC}, which leverages the problem characteristics. The introduced dynamics are not only in full corroboration with the color constancy theory, i.e., Eq.~\eqref{eq1}, but they also enable us to substantially reduce the number of the required parameters by up to 90\% while achieving comparable experimental results to previous state-of-the-art approaches.  This makes our approach energy and time-efficient and, thus, suitable for low-cost devices.  
  
Furthermore, illuminant  estimation approaches typically fail for some samples and yield very high errors. For the practical use of these techniques, it would be important to be able to identify these extreme error cases to know when the model prediction for a given scene is not reliable and a different algorithm should be preferred. This can be seen as an uncertainty estimation problem. Recently, Monte Carlo dropout was proposed in the illuminant  estimation context to predict the model uncertainty \cite{laakom2020monte}. However, this requires multiple forward passes of the same image to produce an uncertainty estimate which is problematic as it further increases the energy and time costs of the process. In this work, we propose a novel computational color constancy uncertainty estimation approach that augments the trained model with an auxiliary branch that learns to predict the error based on the feature representation. Intuitively, the model learns which feature combinations are robust and, thus, likely to yield low errors and which combinations result in erroneous estimates. Our approach is efficient as it requires only a single forward pass of the input. 

To summarize, our main contributions are as follows: 

\begin{itemize}
\item We propose a channel-wise decomposition of the illuminant  estimation problem into three independent sub-problems.  

\item We propose a novel CNN-based illuminant estimation approach, called \acf{CWCC}, which leverages the decomposition enabling us to reduce the number of parameters up to 90\% 

\item We propose a novel efficient uncertainty estimation approach that augments the trained model with an auxiliary branch that learns to predict the error based on the feature representation. 
\end{itemize}

\section{Channel-wise color constancy}
We note that the problem defined in Eq.~\eqref{eq1} can be divided into three problems using the color channels (r,g,b):
\begin{equation}  \label{eq2}
 \textbf{I}_r =  \textbf{R}_r   \textbf{e}_r, \;\;\;\;
  \textbf{I}_g =  \textbf{R}_g   \textbf{e}_g, \;\;\;\;
   \textbf{I}_b =  \textbf{R}_b   \textbf{e}_b.
\end{equation}
As it can be seen, the system composed of the sub-equations of  Eq.~\eqref{eq2} is equivalent to the problem defined in Eq.~\eqref{eq1}. Moreover, we note that the sub-equations in Eq.~\eqref{eq2} are symmetric, i.e., the problem defined in each equation is similar. Typically, CNN-based methods operate directly on the input image and optimize the filters of the first layers jointly without exploiting this symmetry in Eq.~\eqref{eq1}. In this paper, we  argue that this might not be optimal, because it can lead to learning some non-intuitive cross-channel correlations from the scarce training data. Thus, exploiting the symmetry can improve the performance of the CNN-based approaches for illuminant  estimation. To this end, we propose a channel-wise CNN, which solves the sub-equation in Eq.~\eqref{eq2} disjointly. Moreover, we note that our formulation of the illuminant estimation task lifts the non-linearly of the problem. In fact, as it can be seen, in the sub-equations of Eq.~\eqref{eq2},  the connection between the inputs $\textbf{I}_i$ and the unknown $\textbf{e}_i$ is linear as opposite to the standard formation which contains a non-linear operator, i.e., element-wise multiplication.

\begin{figure}[h]
\includegraphics[width=0.45\textwidth]{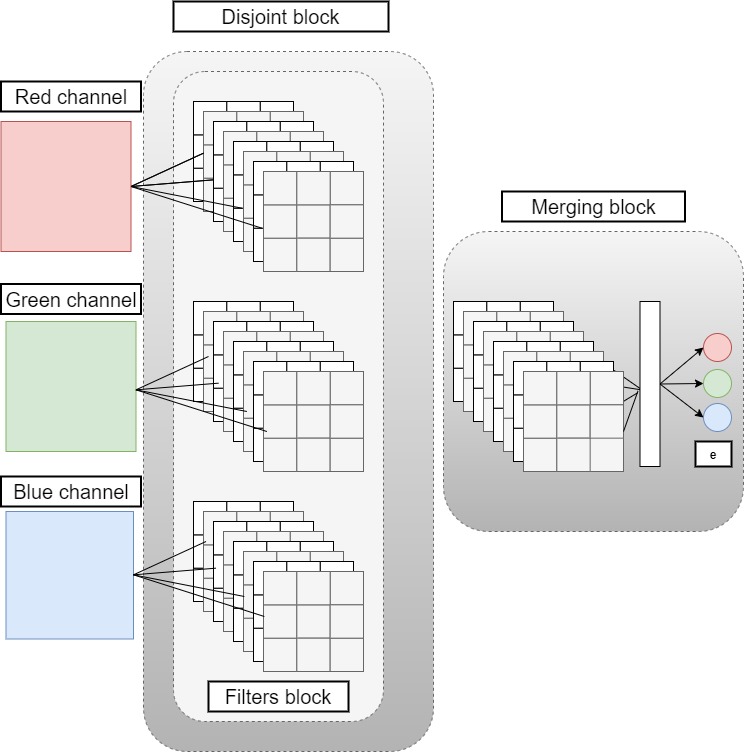}
\centering
 \caption{The full illustration of the proposed approach}
 \label{figure_model}

\end{figure}

The proposed model is presented in Figure \ref{figure_model}. It is composed of two blocks, the disjoint block and the merging block. The disjoint block learns to solve each sub-equation separately. To this end, each color channel has a separate CNN sub-model. Moreover, we exploit the symmetry of the sub-problems by sharing the weights of  'filter blocks' of the three sub-models.  In the merging block,  we concatenate the outputs of each channel of the first block. Then, we use a model which acts on this mixed representation and aims to learn the optimal way to merge the feature maps of each channel and approximate the illuminant $\textbf{e}$.

Formally, given an image $ \textbf{I} = [ \textbf{I}_r, \textbf{I}_g,  \textbf{I}_b] $, the outputs of the disjoint block are obtained as follows:
\begin{equation}  \label{eq1_2}
 \textbf{F}_r =   f^1_{\theta}(\textbf{I}_r), \;\;\;\;
  \textbf{F}_g = f^1_{\theta}(\textbf{I}_g),\;\;\;\;
   \textbf{F}_b =  f^1_{\theta}(\textbf{I}_b),
\end{equation}
where $f^1_{\theta}$ is a fully convolutional CNN-model with parameters $\theta$.  $\textbf{F}_r$, $\textbf{F}_g$, and $\textbf{F}_b$ are the output feature maps for the red, green, blue channels, respectively. Using the same model $f^1_{\theta}(\cdot)$ over all the three channels reduces the number of trainable parameters of the total model. Intuitively, the model learns a feature extractor which is independent of the color channel. This constraint is inspired by channel-wise symmetry of the illuminant estimation task. Next, the illuminant estimate is computed as follows: 
\begin{equation} \label{eq1_4}
   \textbf{e} =  f^2_{\theta'}(GAP(\textbf{F}_r \frown\textbf{F}_g \frown \textbf{F}_b)),
\end{equation}
where $\frown$ denotes the concatenation of the feature maps of the different channels, $GAP$ is the global average pooling operator compiling a global feature representation of the input image. Finally, the predicted illuminant is obtained through a second model, namely $f^2_{\theta'}$, with parameters ${\theta'}$.  This model takes a vector-representation of the scene and learns to approximate the illuminant and is formed of fully connected layers. The parameters $\theta$ and $\theta'$ of the inner models $f^1$ and $f^2$ can be jointly optimized in an end-to-end matter during the back-propagation. 

For $ f^1_{\theta}$, we use a SqueezeNet-like \cite{iandola2016squeezenet} fully convolutional architecture composed as follows: First, we have a convolutional layer with 64 $3\times3$ kernels, then a $3\times3$ maxpooling layer. Next, we use two fire modules \cite{iandola2016squeezenet} with a size of 64 followed by a $3\times3$ maxpooling layer. At the end, we have two fire modules with a size of 128 followed by a $3\times3$ maxpooling layer. For the second inner-model of the merging block $f^2_{\theta'}$ we use a fully connected model containing a fully connected layer with 40 units and ReLu activation and a dropout regularizer with rate of 10\%. The output of this layer is connected to the output layer composed of 3 units.

\section{Uncertainty estimation block}

For the practical use of illuminant estimation techniques, it is important to be able to identify  when the model will fail and when its prediction for a given scene is not reliable. This can be seen as an uncertainty estimation problem \cite{gal2016dropout, loquercio2020general}.
We propose to augment our trained illuminant estimation model to predict the model uncertainty. We add an additional branch linked to the last intermediate layer which aims to learn to predict the error based on the feature representation. Intuitively, the model learns which feature combinations are robust and are thus likely to yield low errors and which combinations result in erroneous estimates.  The predicted error can be seen as an uncertainty estimate as it directly quantifies to expected loss. Similar to an uncertainty measure, it is expected to have high values in the case of high errors and lower values in the case of low errors.  Compared to the existing uncertainty estimation approaches in color constancy  \cite{laakom2020monte}, we note that our approach requires only a single forward pass of the same image to produce an uncertainty estimate, which enables us to save time and energy. The proposed approach can be incorporated also inside any other illuminant estimation method to measure uncertainty. 

\begin{figure}[b]
\includegraphics[width=0.45\textwidth]{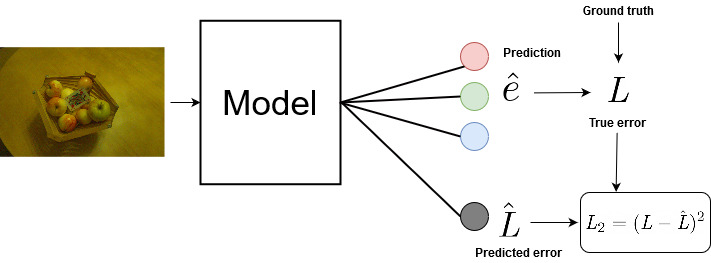}
\centering
 \caption{The full illustration of the proposed uncertainty estimation  scheme}
  \label{uncertainty_fig}

\end{figure}

The full illustration of the proposed scheme is presented in Figure \ref{uncertainty_fig}. Given an input image, we generate two outputs: the main illuminant prediction and the predicted error using an auxiliary branch. As we have access to the ground-truth illuminations of our training samples, we can construct a training set for the additional branch by computing the true errors obtained by the trained illuminant estimation model. While training the uncertainty estimation block, we freeze the prediction part of the network to ensure a 'fixed' representation of every input sample and fine-tune only the additional branch of the network. As the topology of this additional model, we use two fully connected layers with Relu activation of sizes 40 and 15, respectively, and one-dimensional fully connected output layer. This additional model is trained using the mean square error to approximate the error, namely the Recovery error $error_{recovery}$.

\section{Experimental evaluation}

\subsection{Experiments on channel-wise color constancy}

In this section, we validate the performance of our proposed approach empirically. To this end, we use INTEL-TAU dataset \cite{laakom2019intel}, which is the largest publicly available dataset for computational color constancy with 7022 total images split to 10 folds. As in prior works  with this dataset \cite{laakom2020monte,afifi2020cross,laakom2020probabilistic}, the models are evaluated using 10-fold cross validation and the average performance is reported. 

As error metric, we use the Recovery angular error  $error_{recovery}$ \cite{21} which measures the cosine similarity between the ground truth and prediction:
\begin{equation}
       \text{$error_{recovery}$}(\textbf{e}^{gt},\textbf{e}^{est})= \cos^{-1} \left({ \frac{ \textbf{e}^{gt} \textbf{e}^{est}}{\| \textbf{e}^{gt} \| \|\textbf{e}^{est} \| } } \right) 
\end{equation}
For more quantitative insights, we also use the Reproduction angular error $e_{reproduction}$ \cite{finlayson2014reproduction} defined as follows:
\begin{equation}
     \text{$error_{reproduction}$}(\textbf{e}^{gt},\textbf{e}^{est})= \cos^{-1} \left({ \frac{ r(\textbf{e}^{gt} , \textbf{e}^{est})  \hspace{2mm} \textbf{o}}{\| r(\textbf{e}^{gt} , \textbf{e}^{est}) \| } } \right), 
\end{equation}
where $\textbf{e}^{gt}$ is the ground truth illumination, $\textbf{e}^{est}$ is the estimated illumination, $r(\textbf{e}^{gt} , \textbf{e}^{est}) = \textbf{e}^{gt}/ \textbf{e}^{est}$ is the element-wise division of $\textbf{e}^{gt}$ by $\textbf{e}^{est}$, and $\textbf{o}$ is the normalized unit vector, i.e., $\textbf{n} = [1/\sqrt3, 1/\sqrt3, 1/\sqrt3 ]^T$.  For both error metrics, we report the average of the best 25\%, the average, the median, the trimean, and the average of the worst 25\%  of the test errors.

\begin{table*}[t]
	\caption{Results of benchmark methods on INTEL-TAU dataset using cross-validation protocol.}
		\label{inteltauuncer}
 \centering	
\scriptsize
\begin{tabular}{l|llllc||llllc}
                    & \multicolumn{5}{c||}{$error_{recovery}$}                                                       & \multicolumn{5}{c}{$error_{reproduction}$}                                                  \\ \hline
Method              & Best \newline 25\% & Mean & Med. & Tri. & W. \newline 25\% & Best \newline 25\% & Mean & Med. & Tri. & W. \newline 25\% \\
\hline
Grey-World  & 1.0   & 4.9  & 3.9  & 4.1  & 10.5  & 1.2   & 6.1  & 4.9  & 5.2  & 13.0    \\
White-Patch     & 1.4  & 9.4  & 9.1  & 9.2  & 17.6  & 1.8   & 10.0  & 9.5  & 9.8  & 19.2 \\
Grey Edge    & 1.0  & 5.9  & 4.0  & 4.6  & 13.8& 1.2  & 6.8  & 4.9  & 5.5  & 13.5  \\
Grey Edge 2 & 1.0  & 6.0  & 3.9  & 4.8  & 14.0 & 1.2   & 6.9  & 4.9  & 5.6  & 15.7 \\
Shades-of-Grey  & 0.9 & 5.2  & 3.8  & 4.3  & 11.9   & 1.1  & 6.3  & 4.7  & 5.1  & 13.9   \\
Cheng et al. 2014 & 0.7 & 4.5  & 3.2 & 3.5 & 10.6 &  0.9 & 5.5  & 4.0  & 4.4  & 12.7 \\
Weighted GE   &  0.8& 6.1  & 3.7  & 4.6  & 15.1& 1.1 & 6.9  & 4.5  & 5.4  & 16.5   \\
Yang et al. 2015  &  0.6                               &  3.2  &  2.2  &  2.4  &  7.6    &  0.7                               &  4.1  &  2.7  &  3.1  &  9.6                            \\
Color Tiger           & 1.0     & 4.2    & 2.6  & 3.2  & 9.9    & 1.1      & 5.3  &  3.3  & 4.1  & 12.7   \\
Greyness Index  &  0.5                               &  3.9    &  2.3  &  2.7  &  9.8                             &  0.6                               &  4.9  &  3.0  &  3.5  &  12.1   \\

PCC\_Q2  &  0.6                               &  3.9    &  2.4  &  2.8  &  9.6                             &  0.7 &  5.1& 3.5  &  4.0  &  11.9  \\
\hline

FFCC &  0.4   &  2.4    & 1.6  & 1.8  & 5.6 &  0.5   & 3.0  & 2.1  & 2.3  & 7.1 \\

Bianco &  0.9&  3.5   & 2.6  & 2.8  & 7.4                             &  1.1                            & 4.4 & 3.4  & 3.6  & 9.4 \\
C3AE&  0.9&  3.4    & 2.7  & 2.8  & 7.0                             &  1.1& 3.9  & 3.3  & 3.5  & 8.8 \\

BoCF &  0.7&  2.4   & 1.9  & 2.0  & 5.1                           &  0.8    & 3.0& 2.3  &  2.5& 6.5 \\
FC4 (VGG16) &  0.6&  2.2    & 1.7  & 1.8  & 4.7  &  0.7 & 2.9  & 2.2  & 2.3  & 6.1 \\

\hline
\ac{CWCC}  &  0.7& 2.4   & 1.9  & 2.0  & 4.9 &  0.8  & 3.0 & 2.3  & 2.7 & 6.3 \\

\end{tabular}
\end{table*}
In Table \ref{inteltauuncer}, we provide the results for the following unsupervised approaches: White-Patch \cite{d5}, Grey-World \cite{d4}, Color-PCA \cite{nus}, Shades-of-Grey \cite{d6}, Weighted Grey-Edge \cite{d8}, Greyness Index 2019  \cite{qian2019finding}, Color Tiger   \cite{banic2017unsupervised}, PCC\_Q2 \cite{laakom2020probabilistic}, and the method reported in \cite{yang2015efficient}.  For the supervised approaches, we report the results of  Fast Fourier Color Constancy  (FFCC) \cite{46440} and the following five \ac{CNN}-based approaches:  \ac{FC4} \cite{44}, Bianco \cite{22}, C3AE \cite{mine}, and BoCF \cite{f1} along with our approach \ac{CWCC}.

As can be seen, the supervised approaches consistently outperform the unsupervised methods across all metrics. This is due to the fact that unsupervised approaches typically rely on strong assumptions regarding the content of the scene. Thus, when these assumptions are violated, the methods fail. On the contrary, to learning-based approaches where the illuminant estimation is learned end-to-end without any prior assumptions lead to better performance. 

\begin{figure*}[h]
\includegraphics[width=0.9\textwidth]{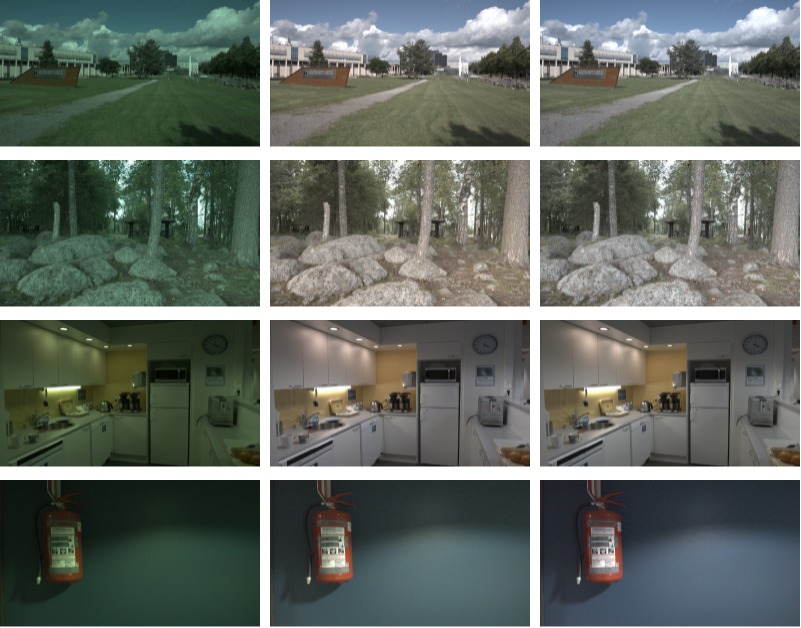}
\centering
 \caption{Visual results on four samples of INTEL-TAU. From left to right: Input image, \ac{CWCC}  output, and ground truth image. The corresponding $error_{recovery}$ errors from top to down are: 2.95, 3.53, 1.41, and 5.92.  }
  \label{visualization}

\end{figure*}

The proposed approach \ac{CWCC} outperforms the state-of-the-art unsupervised approaches. For example, in terms of the worst 25\%, \ac{CWCC} yields better results compared to the method in Yang et al. 2015 by $2.7^{\circ}$. Compared to learning-based approaches,  we note that  \ac{CWCC} outperforms Bianco and C3AE across all metrics and achieves comparable results compared to FFCC  and FC4. In fact compared to FFCC, which is not a CNN-based approach our methods performs better in the extreme cases. This is clear in terms of the worst 25\% metric where our approach yields $0.7^{\circ}$ and $0.8^{\circ}$  improvement in the Recovery and the Reproduction errors, respectively. Compared to the CNN-based approach FC4, we note that our achieves competitive results across all the metrics while using less than 10\% of the number of parameters, as illustrated in Table \ref{num_parameters}. This supports our assumptions that channel-wise decomposition is reasonable in the illuminant estimation context. 
\begin{table}[t]
\renewcommand{\arraystretch}{1}
 \centering	
 %\footnotesize\setlength{\tabcolsep}{2pt}
	\caption{Number of parameters of different \ac{CNN}-based approaches}
	\label{num_parameters}
	\begin{tabular}{l|r} % four columns, alignment for each
		\hline
Method         &  \# parameters  \\	
\hline
Bianco   &   154k   \\
Fc4(SqueezeNet)    &   1.9M  \\
FC4 (AlexNet) & 3.8M \\
DS-Net&   17.3M   \\
%BoCF(2conv+150 words + attention1) &  376k \\
\hline
\ac{CWCC} &  155k \\
	\end{tabular}
\end{table}
Figure \ref{visualization} illustrates some visual result of our approach on four different test samples from INTEL-TAU. As it can be seen, our approach generalizes well for different environments.

To further illustrate the usefulness of exploiting the symmetry between the sub-equations of our problem formulation in Eq.~\eqref{eq2} via weight sharing, we perform an ablation study by comparing the performance of our method \ac{CWCC} to a variant of our model, called \ac{CWCC}*, that uses a different feature extractor for each channel. Formally, the shared model $f^1_{\theta}$ in Eq.~\eqref{eq1_2} is replaced as follows:
\begin{equation}  \label{eq1_2ablation}
 \textbf{F}_r =   f^1_{\theta_r}(\textbf{I}_r), \;\;\;\;
  \textbf{F}_g = f^1_{\theta_g}(\textbf{I}_g),\;\;\;\;
   \textbf{F}_b =  f^1_{\theta_b}(\textbf{I}_b),
\end{equation}

The empirical result of this model on INTEL-TAU are presented in Table \ref{inteltau_abilation}. As can be seen, even though removing the weight sharing constraint gives the model more flexibility, the performance of the model declines in all metrics. This is clear in terms of the worst 25\%, where \ac{CWCC}* yields a higher error by $0.5^{\circ}$ compared to the standard \ac{CWCC}. This can be explained by the data scarcity. In fact, removing the weight sharing constraint almost triples the number of trainable parameters to be optimized. As the training data is limited, this leads \ac{CWCC}* to overfit to the training data and to fail to generalize well for the unseen test samples. 

\begin{table*}[h]
	\caption{Ablation study results of \ac{CWCC} on INTEL-TAU dataset using cross-validation protocol.}
		\label{inteltau_abilation}
 \centering	
\scriptsize
\begin{tabular}{l|llllc||llllc}
                    & \multicolumn{5}{c||}{$error_{recovery}$}                                                       & \multicolumn{5}{c}{$error_{reproduction}$}                                                  \\ \hline
Method              & Best \newline 25\% & Mean & Med. & Tri. & W. \newline 25\% & Best \newline 25\% & Mean & Med. & Tri. & W. \newline 25\% \\
\hline
\ac{CWCC}  &  0.7& 2.4   & 1.9  & 2.0  & 4.9 &  0.8  & 3.0 & 2.3  & 2.7 & 6.3 \\

\ac{CWCC}*  &  0.8& 2.7   & 2.1  & 2.2  & 5.4 &  0.9  & 3.3 & 2.6  & 2.8 & 6.9 \\
\end{tabular}
\end{table*}

\subsection{Experiments on uncertainty estimation}

\begin{figure*}[h]
\includegraphics[width=\textwidth]{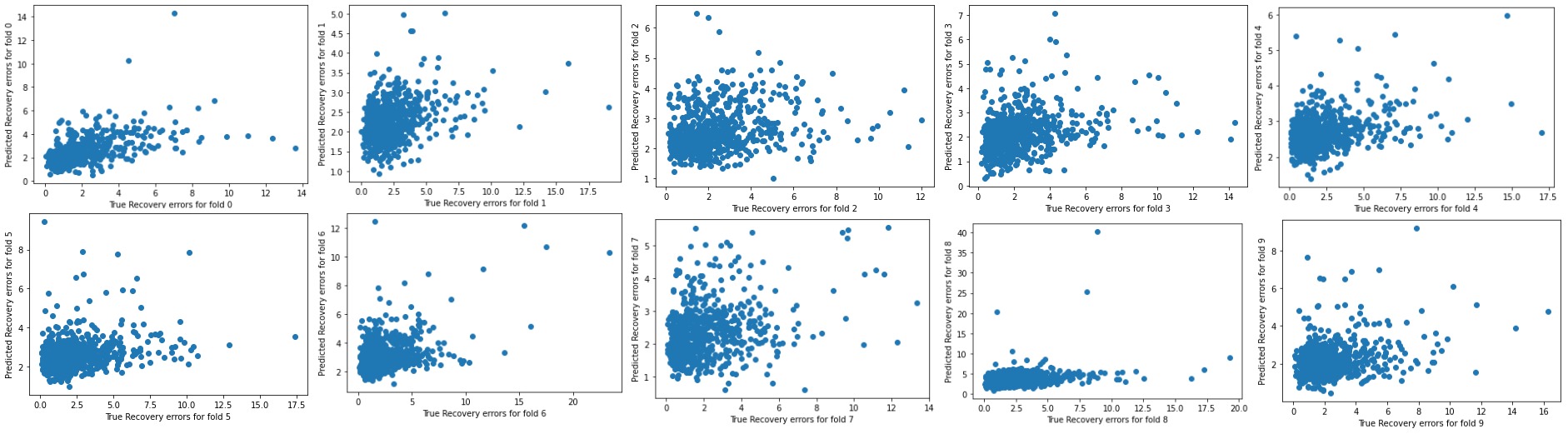}
\centering
 \caption{Predicted loss vs true loss using the proposed approach on the different folds of INTEL-TAU. The correlation coefficients from fold 0 to 10 are: 0.47, 0.34, 0.24, 0.25, 0.34, 0.30, 0.45, 0.28, 0.33, and 0.31. }
  \label{uncertainty_plot}

\end{figure*}

The results for uncertainty estimation on the test samples of the different INTEL-TAU folds are presented in Figure \ref{uncertainty_plot}. For most of the samples, the predicted error correlates well with the true error and the model is able to tell how confident it is about its illuminant prediction. However, it is worth noting that for some extreme examples, the model is over-confident, i.e., predicting low error even though the true errors are high ($>6^{\circ}$). Eliminating these cases can be achieved by setting a lower threshold on the predicted loss. For example, we could set our threshold on when to rely on the model at $2.5^{\circ}$ predicted error. Then, we can guarantee with a high probability that the true errors will not exceed $5^{\circ}$. However, using this strategy can lead to many false negatives, i.e., deciding not to rely on the model because the predicted uncertainty is higher than the threshold while the model prediction is actually good. So there is a trade-off to be made depending on the application.  

\section{Conclusion}

In this paper, we proposed a channel-wise decomposition of the color constancy problem into three independent sub-problems. Based on this decomposition, we proposed a novel CNN-based illuminant estimation approach, called \acf{CWCC}, which leverages this formulation. The proposed method substantially reduces the number of parameters needed  to solve the task by up to 90\% while achieving competitive experimental results compared to state-of-the-art methods. Moreover, we proposed a novel efficient color constancy uncertainty estimation approach that augments the trained model with an auxiliary branch that learns to predict the error based on the feature representation. We showed empirically that the proposed technique can indeed be used to avoid several extreme error cases and, thus, improves the practicality of the proposed technique.  Our future research directions include improving the uncertainty estimation approach to generalize better on the extreme cases.

\ifCLASSOPTIONcompsoc
  \section*{Acknowledgments}
\else
  \section*{Acknowledgment}
\fi

This  work  has been  supported  by  the NSF-Business  Finland Center for Visual and Decision Informatics (CVDI) project AMALIA. The work of Jenni Raitoharju was funded by the Academy of Finland (project 324475).

\ifCLASSOPTIONcaptionsoff
  \newpage
\fi

\bibliographystyle{IEEEtran}

\bibliography{strings}

\end{document}